\documentstyle[epsfig,aps,fleqn,preprint]{revtex}
\draft
\begin{document}
\title{Phase Matching in Quantum Searching}
\author{Gui Lu Long\thanks{Corresponding author, Department of Physics,
Tsinghua University, Beijing, P.R. China, Tel 86-10-62782163, Fax
86-10-62781604, email: gllong@mail.tsinghua.edu.cn.}, Yan Song Li, Wei Lin Zhang and Li Niu}
\address{Department of Physics, Tsinghua University, Beijing, 100084, P.
R. China}
\date{May 15, 1999}
\maketitle
\begin{abstract}
Each iteration in Grover's original quantum search algorithm contains 
4 steps: two Hadamard-Walsh transformations and two amplitudes
inversions. When the inversion of the marked state is 
replaced by arbitrary phase
rotation $\theta$ and the inversion for the prepared state
$|\gamma\rangle$ is replaced by rotation through $\phi$, 
we found that these phase rotations must satisfy a matching
condition $\theta=\phi$. 
Approximate formula for the amplitude of the
marked state after an arbitrary number of iterations are also derived. 
When phase
matching is obtained, we found that the generalized quantum search is
still a rotation in 2-dimensional space, 
but with a small angle $\beta'\approx
(2\sin{\theta \over 2})\beta$, where $\theta$ is the angle of phase
rotation of the marked state and $2\beta$ is the angle of U(2) rotation
when the phase rotations are the inversions in the original version. 
We give also a
simple explanation of the phase matching requirement.
\end{abstract}


Shor's prime factoring algorithm and Grover's quantum search algorithm 
are two of the great quantum algorithms\cite{r1,r2}. 
In Grover's original version, the algorithm is composed by: 1) inversion
of the amplitude of the marked state; 2) inversion about the average.
This second step can be constructed by two Hadamard-Walsh
transformations and the inversion of the amplitudes of all basis states
different from $|0\rangle$.  There are many applications of the
algorithm. For example, using the algorithm, one can search a large
database using only a single query\cite{r3}. It can be used in
deciphering the DES encryption scheme\cite{r4}. It can  be
used in quantum counting\cite{r5}. 
As Grover's algorithm involves only simple operations, it
is easy to implement in experiment. It has been successfully implemented
in NMR techniques\cite{r6,r7} together with the simple D-J
problem\cite{r8,r9}.

Grover's algorithm is optimal. According to Benett et al \cite{r10},
no quantum algorithm can solve the search better than $O(\sqrt{N})$.
Boyer et al \cite{r11} have given analytical expressions for the
amplitude of the states in Grover's search algorithm and  tight
bounds on the algorithm. Zalka\cite{r12} has improved this tight bounds
and showed that Grover's algorithm is optimal. Zalka
also proposed\cite{r13} an improvement on Grover's algorithm. 
Biron et al\cite{r14} generalized Grover's algorithm to an
arbitrarily distributed initial state. Pati\cite{r15} recast the
algorithm in geometric language and studied the bounds on the algorithm.
Very recently, Ozhigov\cite{r16} 
showed that quantum search can be further speeded up
by a factor of $\sqrt{2}$ by parallelism.  Gingrich et al\cite{r17} also
generalized Grover's algorithm with parallelism with improvement.
Jozsa\cite{r18} gave a simple
explanation of Grover's quantum search algorithm in simple geometry.

To generalize Grover's quantum algorithm, one can do: 1) replace the
Hadamard-Walsh algorithm by an arbitrary unitary transformation. This
has been done neatly by Grover\cite{r19}; 2) one can replace the
inversions of the amplitudes by arbitrary phase rotations as 
suggested in
the same Letter by Grover\cite{r19}. It is believed that such a
replacement will lead to a quantum search algorithm, though not as
fast as the original version. Some authors have used this generalization
to the advantage to produce a small iteration in a quantum search
algorithm such that the amplitude of the marked state will be exactly 
unity\cite{r5,r13}, thus avoiding the ``over-cooking'' problem in
quantum searching. For instance, Zalka proposed to rotate the
phase of the marked state by a small angle rather than $\pi$ to produce
a smaller rotation of the state vector of the quantum
computer\cite{r13}. However, it is found recently that
an arbitrary phase(except $(2i+1)\pi$, where $i$ is an integer)
rotation of the marked state together with the inversion about average
can not do any quantum search at all\cite{r20}, contrary to what      
expected.

Grover has shown that in general $I_{\gamma}=I-2|\gamma\rangle\langle
\gamma|$ and $I_{\tau}=I-2|\tau\rangle\langle\tau|$ together with an
arbitrary unitary operator $U$ can be used to construct a quantum search
algorithm $Q=-I_\gamma U^{-1}I_\tau U$.  When $U^{-1}=U=W$, the
Hadamard-Walsh transformation(denoted by $W$ hereafter), 
and $|\gamma\rangle=|0\rangle$, one
recovers the original Grover's quantum algorithm. It is worth pointing
out that in his first paper\cite{r2}, the inversion about average is
realized by $-WI_{n}W$, where $I_{n}$ is the operation that inverses the
amplitudes of the basis vectors $|n\rangle$ for $n\neq 0$. In Ref.
\cite{r19}, it is realized by $-WI_0W$, where $I_0$ is to inverse the
amplitude of the basis state $|0\rangle$. It can be shown easily that
the two differ only by a phase factor. In general, if one rotates the
phases of the basis states $|n\rangle$ except $n\neq \gamma$ through an
angle $\phi$, it is equivalent to rotating the phase of the basis state
$|\gamma\rangle$ through $-\phi$ angle and leaving other 
basis untouched.

In this Letter, 
instead of using inversions which are
phase rotations through angle $\pi$, we use arbitrary phase
rotations separately for $|\tau\rangle$ and $|\gamma\rangle$. Without
confusion in the notations, we also use $I_{\tau}$ and $I_\gamma$ and
$Q$ for these operations:
\begin{eqnarray}
I_\gamma& =&I-\left(- e^{i\theta }+1\right) |\gamma \rangle \langle \gamma |\\
I_\tau& =&I-\left( -e^{i\phi }+1\right) |\tau \rangle \langle \tau |\\
Q &=&-I_\gamma U^{-1}I_\tau U
\end{eqnarray}
When
$\theta=\phi=\pi$, $U=U^{-1}=W$ and 
$|\gamma\rangle=|0\rangle$ we recover the original
Grover's result.

It is easy to calculate that operator $Q$ is represented by the following
matrix in the space span by $|\gamma\rangle$ and $U^{-1}|\tau\rangle$. 
\begin{eqnarray}
Q\left( 
\begin{array}{c}
|\gamma \rangle \\ 
U^{-1}|\tau \rangle
\end{array}
\right) =\left( 
\begin{array}{ll}
-e^{i\theta }-\left(- e^{i\theta }+1\right) \left( -e^{i\phi }
+1\right) \left|
U_{\tau \gamma }\right| ^2 & \left(-e^{i\phi }+1\right) 
U_{\tau \gamma } \\ 
\left( -e^{i\theta }+1\right) e^{i\phi }U_{\tau \gamma }^{*} 
& -e^{i\phi }
\end{array}
\right) \left( 
\begin{array}{c}
|\gamma \rangle \\ 
U^{-1}|\tau \rangle
\end{array}
\right) 
\end{eqnarray}

The basis vectors are not orthogonal. We  orthonormalize them
in a new basis
\begin{eqnarray}
|1\rangle &=&\left( |\gamma >-U_{\tau \gamma }U^{-1}|\tau \rangle \right)
/\xi \\ ,
|2\rangle &=&U^{-1}|\tau \rangle ,
\end{eqnarray}
where $\xi =-\sqrt{1-\left| U_{\tau \gamma }\right| ^2}$. The new
matrix for operator $Q$ is
\begin{eqnarray}
\left( 
\begin{array}{cc}
-e^{i\theta }-\left| U_{\tau \gamma }\right| ^2\left( 1-
e^{i\theta }\right) & 
-\left( 1-e^{i\theta }\right) U_{\tau \gamma }
\sqrt{1-\left| U_{\tau \gamma}\right| ^2} \\ 
-U_{\tau \gamma }^{*}e^{i\phi }\left( 1-e^{i\theta }\right) 
\sqrt{1-\left|
U_{\tau \gamma }\right| ^2} & -e^{i\phi }\left( 1-\left( 1-e^{i\theta
}\right) \left| U_{\tau \gamma }\right| ^2\right)
\end{array}
\right) 
\end{eqnarray}

To study the effect of phase rotations, 
we have studied, by directly multiplying the matrix $Q$ a given number
of times with specified $\theta$ and $\phi$ values. 
For simplicity we put $N=100$,
$|\gamma\rangle=|0\rangle$.  The unitary transformation $U$ is taken as
$W$, the Hadamard-Walsh transformation. We have found 
that there exists a phase matching condition in
constructing a useful quantum search algorithm. This relation is
$\theta=\phi$. 

In Fig.1 and Fig.2, we took $\theta={\pi\over 2}$ and  plotted the
norm of the marked state amplitude after 7 and 8 iterations respectively
for various values of $\phi$. We see the highest peak in both figures
is at $\phi=\theta$. This shows that with phase matching the
probability is the largest. For other values of $\phi$ the probability
are small. When the number of iterations changes, the shape of the
curve changes. For instance, in $|B_7|$ there are 4 peaks between
${\pi\over 2}$ and $\pi$, whereas in $|B_8|$, the number of peaks is 5.
This is not what to be expected for a working quantum search algorithm,
where as $j$ increases the norm of the amplitude should increase
monotonically for small $j$ values.

In Fig.3 and 4, we have plotted $|B_{j}|$ versus $j$ for
$\theta=\phi=\pi/2$, 
$\pi/10$ respectively. 
We see that when phase
matching is satisfied, one can still construct a quantum search algorithm.
The larger the rotating angle, the faster the algorithm. For instance
when $\theta=\phi=\pi/2$, it requires approximately 10 iterations to reach 1.
For $\theta=\phi=\pi/10$, it requires nearly 50
iterations to reach 1. 

In Fig. 5, we have plotted the norm $|B_{j}|$ versus $j$ for a phase
mismatching case: $\theta={\pi\over 2}$ and $\phi={\pi\over 10}$. We see
that the norm of the amplitude of the marked state is restricted to a
narrow range between 0.09 and 0.15. 
Even after 200 iterations,
the amplitude still can not be enhanced. Thus it can
not be used for searching at all.

To see the effect of phase matching on quantum searching algorithms, we
have also plotted the norm of the amplitude of the marked state as a
function of $\phi$ and $\theta$ in Fig.6 for $|B_8|$.  
We see from the 3D plot
clearly the mountain peaks along $\theta=\phi$ 
when the phases are matched. For other
cases,  the values are  far less than 1.

In Fig. 7.,  we have given a 3D plot for $|B_{j}|$ as a function of
$\theta$ and $j$. Here $\phi$ is chosen $\pi/2$. 
We see the $|\sin|$ like behavior of
$|B_{j}|$ as a function of $j$ when phase matching is obtained. 
In other areas, the values of $|B_{j}|$ are small.
It is only with phase
matching that we can construct a quantum search algorithm.

Next, we discuss analytically the situations.
In general, the matrix elements of the unitary transformation can be
written as
\begin{eqnarray}
U_{\tau \gamma }=e^{i\zeta }\sin \left( \beta \right) .
\end{eqnarray}
The matrix of $Q$ can be written in a simple
form
\begin{eqnarray}
Q=\left( 
\begin{array}{cc}
ie^{i\frac \theta 2} & 0 \\ 
0 &-i e^{i\left( \phi +\frac \theta 2\right) }
\end{array}
\right) \left[ i\cos \left( \frac \theta 2\right) \left( 
\begin{array}{cc}
1 & 0 \\ 
0 & -1
\end{array}
\right) -\sin \left( \frac \theta 2\right) \left( 
\begin{array}{cc}
\cos \left( 2\beta \right) & -\sin \left( 2\beta \right) e^{i\zeta } \\ 
\sin \left( 2\beta \right) e^{-i\zeta } & \cos \left( 2\beta \right)
\end{array}
\right) \right].
\end{eqnarray}

In general $\beta$ is small and in the 
order of $1/\sqrt{N}$ where $N$ is the
dimension of the database.
Approximately we can write
\begin{eqnarray}
Q\doteq \left( 
\begin{array}{cc}
ie^{i\frac \theta 2} & 0 \\ 
0 & -ie^{i\left( \phi +\frac \theta 2\right) }
\end{array}
\right) \left( 
\begin{array}{cc}
ie^{i\frac \theta 2} & 2\sin \left( \frac \theta 2\right) e^{i\zeta }\beta \\ 
2\sin \left( \frac \theta 2\right) e^{-i\zeta }\beta & -ie^{-i\frac \theta 2}
\end{array}
\right) 
\end{eqnarray}
 By ignoring an overall 
phase factor $-e^{i {\theta+\phi \over 2}}$, 
we can rewrite the matrix as
\begin{eqnarray}
Q=\left( 
\begin{array}{cc}
e^{i\frac{\theta -\phi }2} & -i2\sin \left( \frac \theta 2\right) \beta
e^{-i\left( \frac \phi 2-\zeta \right) } \\ 
-2i\sin \left( \frac \theta 2\right) \beta e^{i\left( \frac \phi 2-\zeta
\right) } & e^{i\frac{\phi -\theta }2}
\end{array}
\right) 
\end{eqnarray}
Let $\beta ^{\prime }=2\sin \left( \frac \theta 2\right) \beta $,
then
\begin{eqnarray}
Q=\left( 
\begin{array}{cc}
e^{i\frac{\theta -\phi }2} & -i\beta ^{\prime }e^{-i\left( \frac \phi 2-\zeta
\right) } \\ 
-i\beta ^{\prime }e^{i\left( \frac \phi 2-\zeta \right) } & e^{i\frac{\phi
-\theta }2}
\end{array}
\right).
\end{eqnarray}

When the phase matching condition is satisfied, $\theta =\phi $, we have
\begin{eqnarray}
Q_{new}=\left( 
\begin{array}{cc}
1 & -i\beta ^{\prime }e^{-i\left( \frac \phi 2-\zeta \right) } \\ 
-i\beta ^{\prime }e^{i\left( \frac \phi 2-\zeta \right) } & 1
\end{array}
\right)=I+\beta G\approx e^{\beta G}.
\end{eqnarray}
where
\begin{eqnarray}
G=\left( 
\begin{array}{cc}
0 & -ie^{-i\left( \frac \phi 2-\zeta \right) } \\ 
-ie^{i\left( \frac \phi 2-\zeta \right) } & 0
\end{array}
\right) 
\end{eqnarray}
Using the properties of $G$, that is,
\begin{eqnarray}
G^2=-I,
G^3=-G,G^4=I,
\end{eqnarray}
we can obtain for small $\beta$ the following expression for the product
of $Q$,
\begin{eqnarray}
Q^j&=&\cos j\beta ^{\prime }I+\sin j\beta ^{\prime }G=\left( 
\begin{array}{cc}
\cos \left( j\beta ^{\prime }\right) & ie^{-i\left( \frac \phi 2-\zeta
\right) }\sin \left( j\beta ^{\prime }\right) \\ 
ie^{i\left( \frac \phi 2-\zeta \right) }\sin \left( j\beta ^{\prime }\right)
& \cos \left( j\beta ^{\prime }\right)
\end{array}\right) \nonumber\\
&=&\left(\begin{array}{ll} \cos(j\beta') & -\sin(j\beta') \\
			\sin(j\beta') & \cos(j\beta') \end{array}\right),
\label{ekey}
\end{eqnarray}
where in the last step the 
phase factor $-ie^{-i({\phi\over 2}-\zeta)}$ is absorbed in the
definition of the basis vector $|1\rangle$(without loss of generality,
we put this phase factor to 1 henceforth).
Equation (\ref{ekey}) is
simply a rotation in the space span by $|1\rangle$ and $|2\rangle$. Each
operation of $Q$ rotates the state vector of the quantum computer an angle
$\beta ^{\prime }=2\sin \left( \frac \theta
2\right) \beta $ towards the state $U^{-1}|\tau\rangle$. After $j_m$
number of operations, the state vector of the computer is closest to
$U^{-1}|\tau\rangle$, then another operation of $U$ will put the state
of the computer to $|\tau\rangle$, the marked state. Then a measurement
of the state of the computer will yield with near certainty the marked
state. Let's look at the $j_m$, the optimal number of iteration steps. 
Suppose the initial state of the
computer is
$A_0|1\rangle+B_0|2\rangle=\cos\beta_0|1\rangle+\sin\beta_0|2\rangle$.
Then the amplitude of the state $U^{-1}|\tau\rangle$ after $j$
iterations is
\begin{eqnarray}
B_{j}=\sin({j\beta'})\cos\beta_0
+\cos({j\beta'})\sin\beta_0=\sin(j\beta'+\beta_0)
\label{ekey2}
\end{eqnarray} 
When $\sin(j_m\beta'+\beta_0)\approx 1$, we have maximum probability 
for the marked state. In this case,
\begin{eqnarray}
j_m\beta'+\beta_0\approx {\pi\over 2}
\end{eqnarray}
or
\begin{eqnarray}
j_m\sin{\theta\over 2}=({\pi\over 2} -\beta_0)/2\beta\approx 
{({\pi\over 4}-{\beta_0\over 2})\sqrt{N} \over \sin{\theta \over 2}}
\end{eqnarray}
We have plotted the ``exact" numerical results of
$j_m\sin(\theta)/2$ for different values of $N$. It is seen that the
product is nearly a constant, validating our treatment.

We see from the equation (\ref{ekey2}) that when state $|\tau\rangle$,
the marked state and the prepared state 
$|\gamma\rangle$  are
rotated the same angle, we can construct a quantum search
algorithm, but now it requires more steps. However there maybe
cases where this can be used in the advantage, such as those proposed in
Ref. \cite{r5,r13}. In addition, a small rotation angle usually means a
shorter realizing time, for instance in NMR, the angle of rotation is
proportional to the duration of the radio frequency signals, this may be
useful in some problems.

When phase matching is not satisfied, 
$\theta \neq \phi $,  we can give an approximate
formula for small $\beta$,
\begin{eqnarray}
Q^j\approx\left( 
\begin{array}{cc}
e^{ij\delta } & -e^{-i \left( j-1\right) \delta  }
\left( \frac{1-e^{2ij\delta }}{1-e^{2i\delta }}%
\right) \beta' \\ 
e^{-i \left( j-1\right) \delta 
 }\left( \frac{1-e^{2ij\delta }}{1-e^{2i\delta }}\right) \beta' & 
e^{-ij\delta }
\end{array}
\right) ,
\end{eqnarray}
where $\delta
=\left( \theta -\phi \right) /2$. We have omitted the high 
order terms of $\beta $. When acted on an initial state of the form
$A_0|1\rangle+B_0|2\rangle$, the amplitude of the state
$|2\rangle=U^{-1}|\tau\rangle$ becomes
\begin{eqnarray}
B_j=e^{-i \left( j-1\right) \delta   }
\left( \frac{1-e^{2ij\delta }}{1-e^{2i\delta }}\right) \beta'
A_0+ e^{ij\delta} B_0.
\end{eqnarray}
Usually $B_0$ is small(For instance, 
in Grover's original algorithm, the initial
state is $|0\rangle=-\sqrt{N-1 \over N}|1\rangle+\sqrt{1\over
N}|2\rangle$, $B_0=\sqrt{1\over N}$). 
The effectiveness of the algorithm relies heavily on the
first term. However, the factor $\left( \frac{1-e^{2ij\delta }}{1-e^{2i\delta
}}\right)$ plays a crucial role in the phase matching. When $\delta$
approaches zero, this factor approaches $j$. More iterations of the
operation will increase the amplitude.
The factor can be written as a sum
$\sum_{k=0}^{j}e^{ik\delta}$. When $\delta$ is not zero
($\delta\gg \beta'$),   from complex analysis, one sees that there
are serious cancellations in the sum, and this 
makes this factor in an order of 1.
Thus the operation does not increase the amplitude, and the quantum
search algorithm fails.

To summarize, we have generalized Grover's quantum search algorithm to
arbitrary phase rotations. We have found that phase matching between the
phase rotation of the marked state and the prepared state
$|\gamma\rangle$ is crucial in constructing a quantum search algorithm.
Approximate formulas are given for the amplitude of the marked state
after arbitrary number of iterations, and reasons are given to the
failure of the algorithm when phase matching is not satisfied. The phase
matching requirement has set a stringent condition
in the experimental realization
of the algorithm, as an imperfect gate operation may lead to phase mismatching
and thus adversely affect the efficiency of the algorithm.

The authors acknowledge the encouragement of Prof. Haoming Chen.

\begin{figure}
\begin{center}
\epsfig{figure=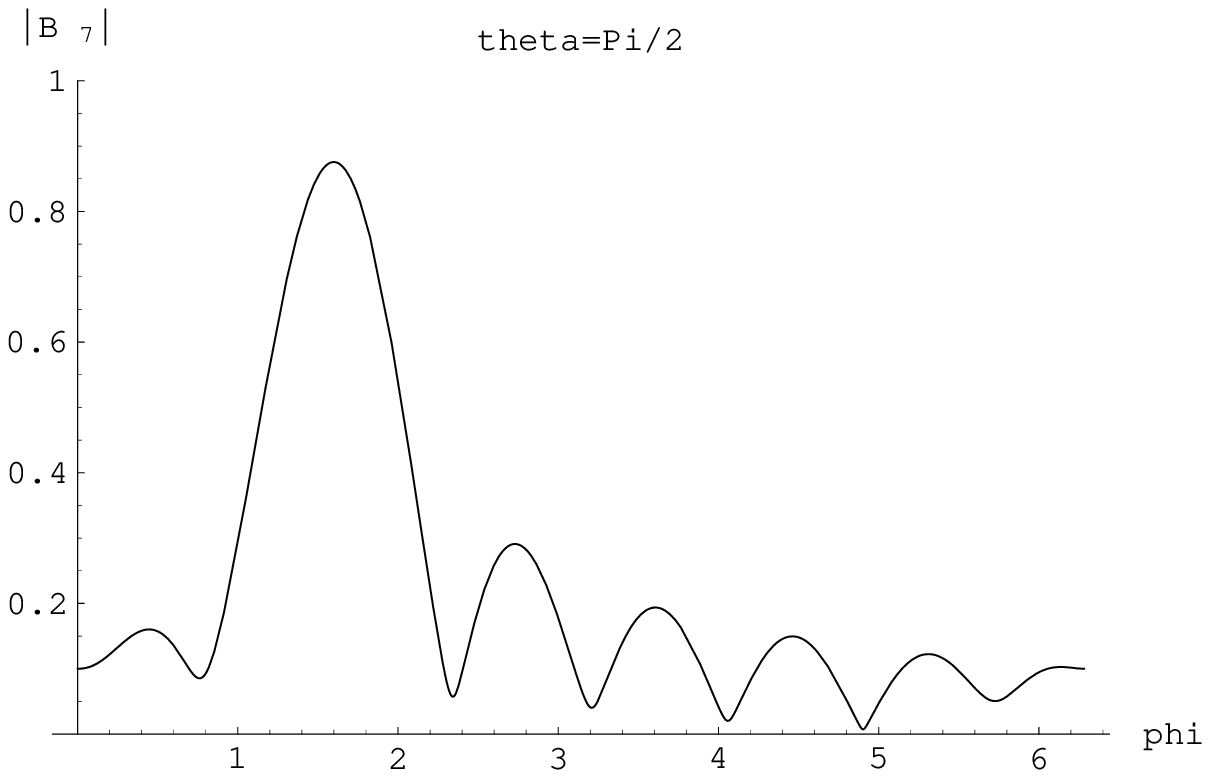,width=10cm}
\end{center}
\caption{$|B_{7}|$ versus $\phi$ for $\theta={\pi \over 2}$.}
\end{figure}

\begin{figure}
\begin{center}
\epsfig{figure=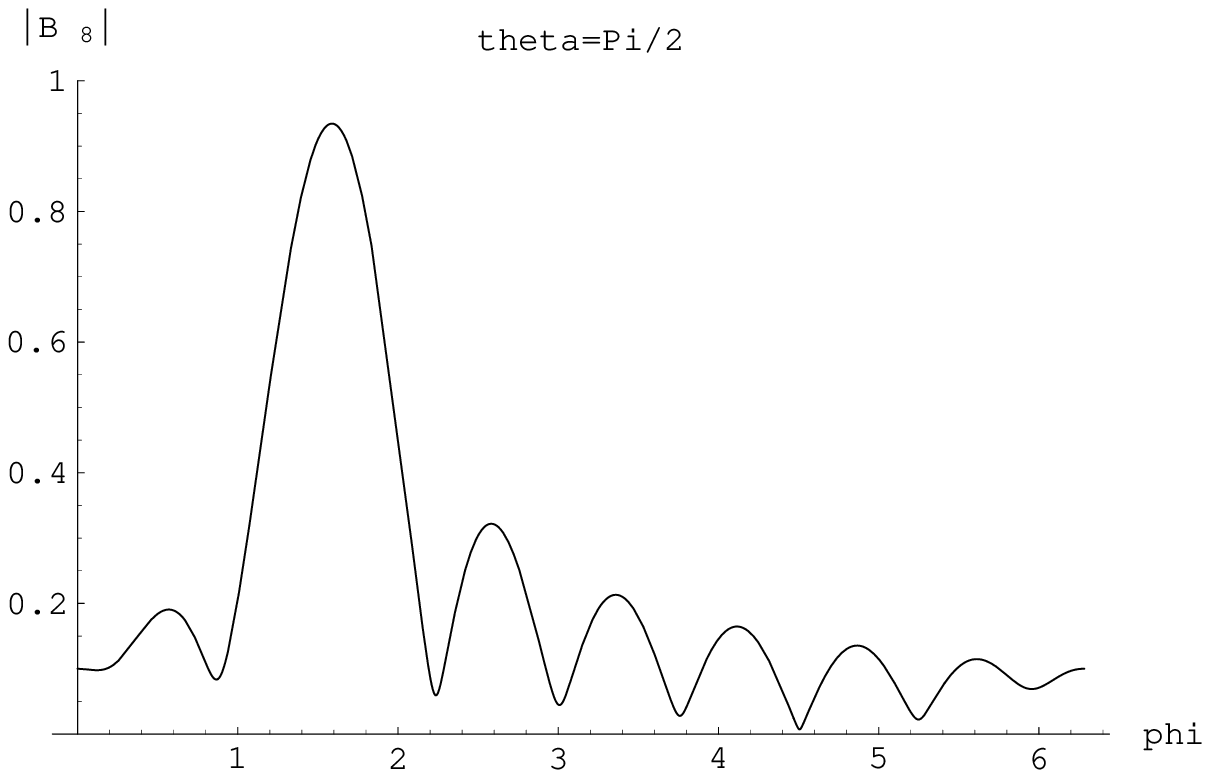,width=10cm}
\end{center}
\caption{$|B_{8}|$ versus $\phi$ for $\theta={\pi \over 2}$.}
\end{figure}

\begin{figure}
\begin{center}
\epsfig{figure=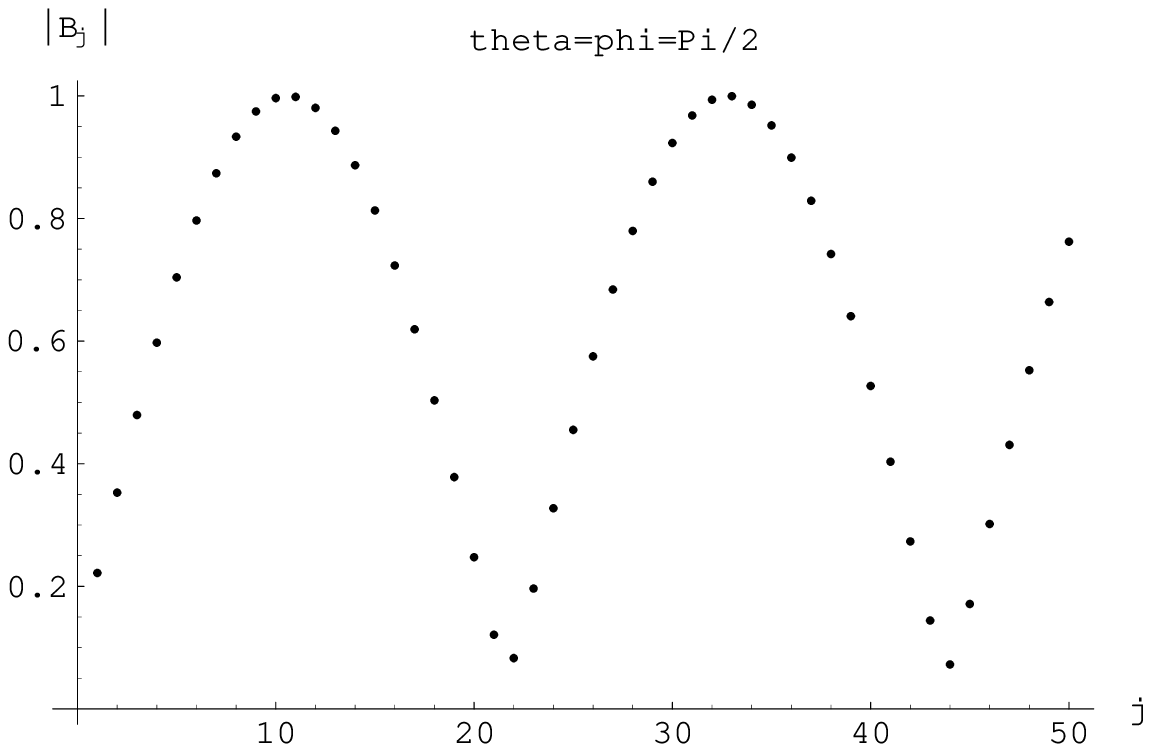,width=10cm}
\end{center}
\caption{$|B_{j}|$ versus $j$ for $\theta=\phi={\pi \over 2}$.}
\end{figure}

\begin{figure}
\begin{center}
\epsfig{figure=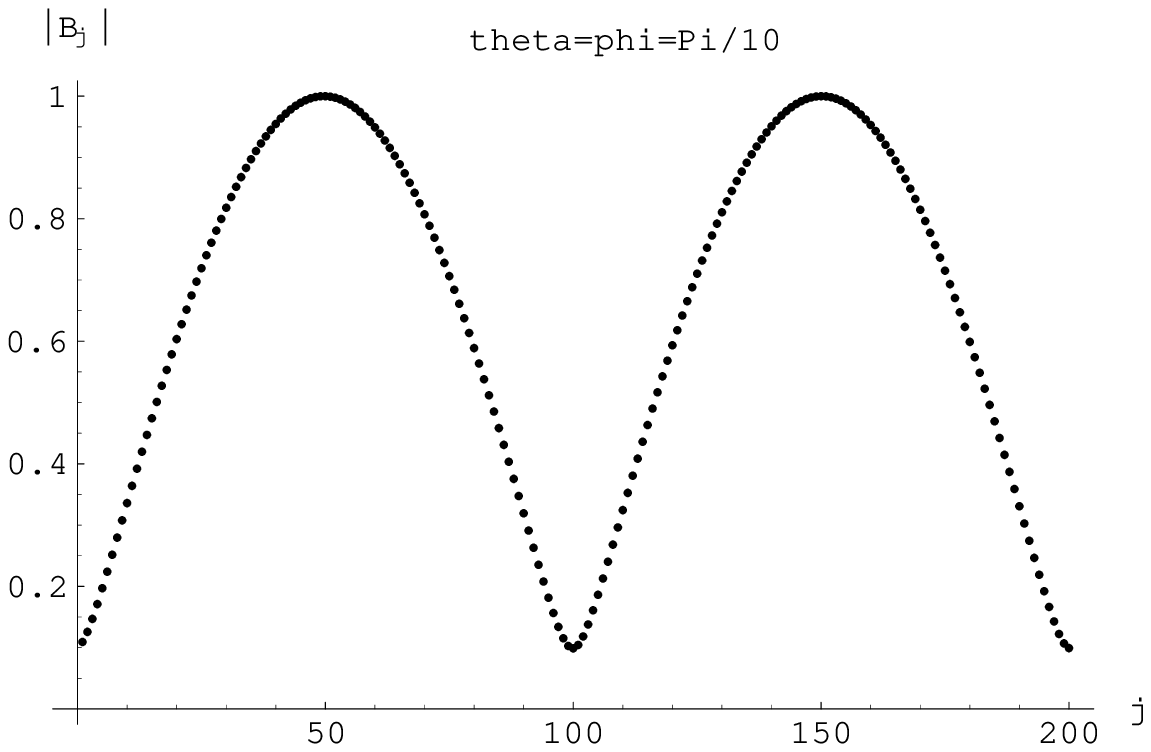,width=10cm}
\end{center}
\caption{$|B_{j}|$ versus $j$ for $\theta=\phi={\pi \over 10}$.}
\end{figure}

\begin{figure}
\begin{center}
\epsfig{figure=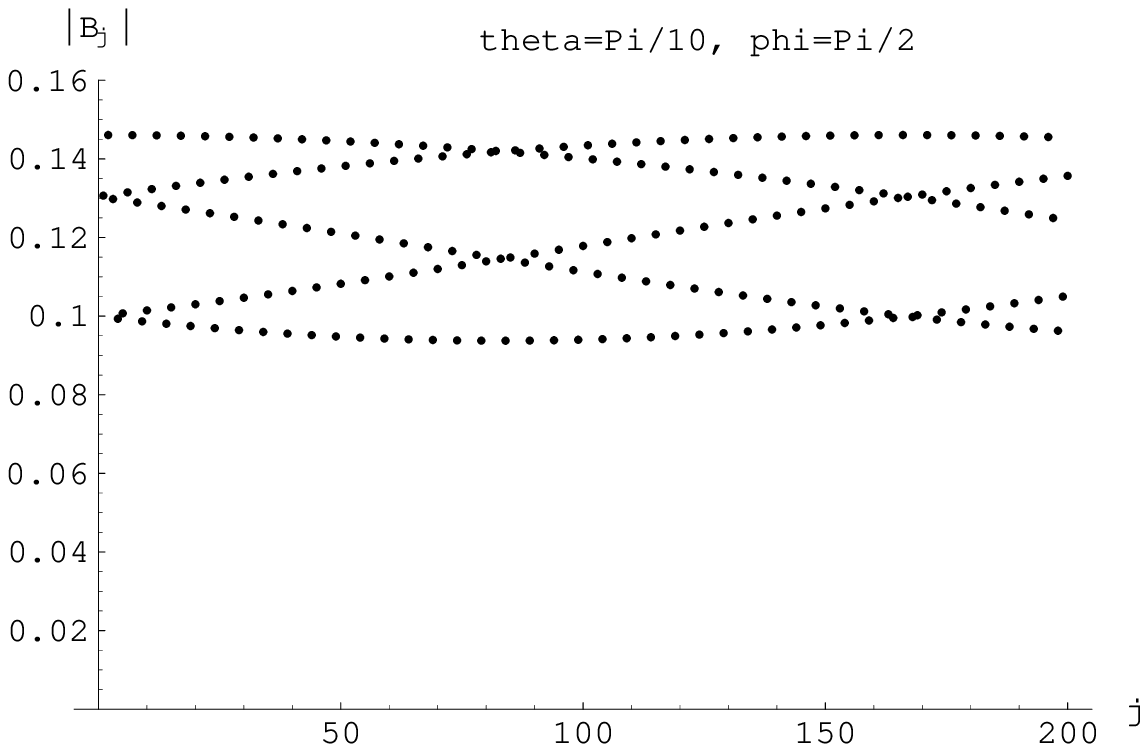,width=10cm}
\end{center}
\caption{$|B_{j}|$ versus $j$ for $\theta={\pi \over 2}$,
$\phi={\pi\over 10}$.}
\end{figure}

\begin{figure}
\begin{center}
\epsfig{figure=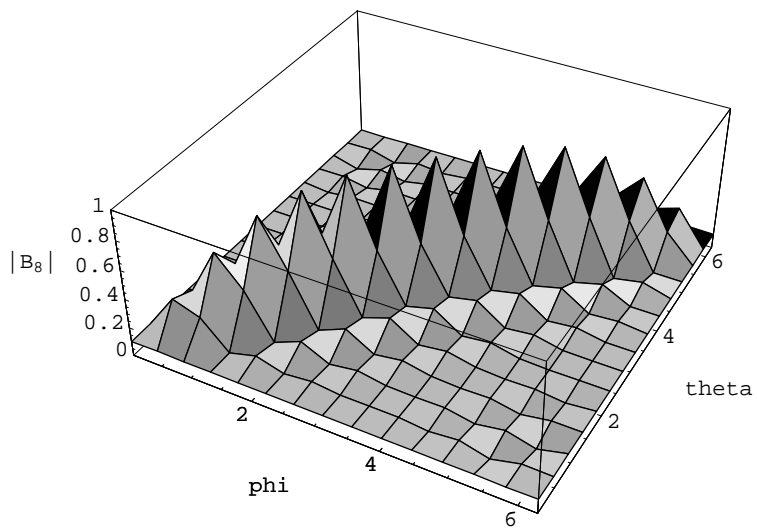,width=10cm}
\end{center}
\caption{3D plot for $|B_{8}|$ versus $\theta$ and $\phi$.}
\end{figure}

\begin{figure}
\begin{center}
\epsfig{figure=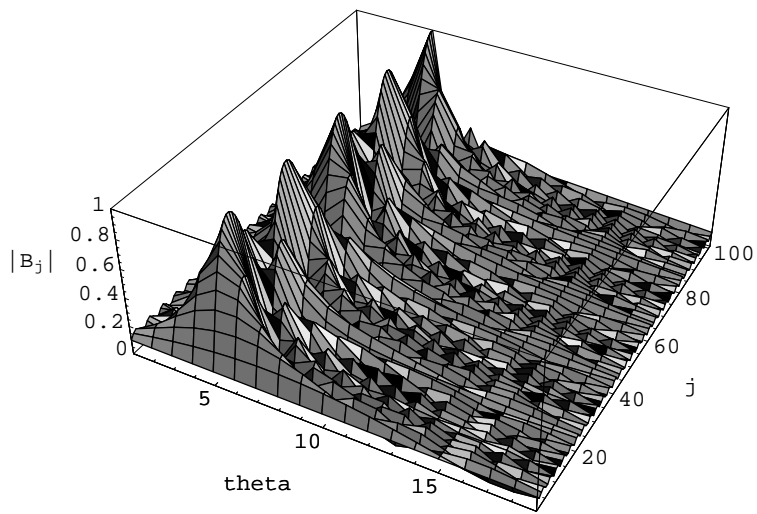,width=10cm}
\end{center}
\caption{3D plot for $|B_{j}|$ versus $\theta$ and $j$. $\phi=\pi/2$ and
$\theta$ is in unit of $\pi/10$.}
\end{figure}

\end{document}